\documentclass[pra,showpacs,twocolumn]{revtex4-1}
\usepackage{amsfonts}
\usepackage{amssymb}
\usepackage{amsmath}
\usepackage{graphicx}
\usepackage{savesym}
\savesymbol{iint}
\usepackage{txfonts}
\restoresymbol{TXF}{iint}
\usepackage{mathrsfs}
\usepackage{bbm}
\usepackage{bm}

\newcommand{\ket}[1]{\left\vert#1\right\rangle}
\newcommand{\bra}[1]{\left\langle#1\right\vert}

\begin{document}

\title{All-optical quantum computing with a hybrid solid-state processing unit}
\author{Pei Pei}
\email{ppei@mail.dlut.edu.cn}
\author{Feng-Yang Zhang}
\author{Chong Li}
\email{lichong@dlut.edu.cn}
\author{He-Shan Song}
\affiliation{School of Physics and Optoelectronic Engineering, Dalian University of Technology, Dalian 116024, P. R. China}
\date{\today}
\begin{abstract}
We develop an architecture of hybrid quantum solid-state processing unit for universal quantum computing. The architecture allows distant and nonidentical solid-state qubits in distinct physical systems to interact and work collaboratively. All the quantum computing procedures are controlled by optical methods using classical fields and cavity QED. Our methods have prominent advantage of the insensitivity to dissipation process benefiting from the virtual excitation of subsystems. Moreover, the QND measurements and state transfer for the solid-state qubits are proposed. The architecture opens promising perspectives for implementing scalable quantum computation in a broader sense that different solid-state systems can merge and be integrated into one quantum processor afterwards.
\end{abstract}
\maketitle

\section{Introduction}
Quantum computer \cite{Nielsen00} is expected to realize the storing, processing and transmitting of quantum information (QI) encoded in many two-level systems (qubits) by coherently controlling the evolution of system following prescribed paths. Quantum computing holds the promise of efficiently solving certain computational tasks intractable by classical algorithms \cite{Shor94,Grover97} and enables the efficient simulation of quantum systems \cite{Lloyd96}. In the past decade, tremendous progress has been made to experimentally implement quantum computing in various physical systems and corresponding coherent control techniques \cite{Ladd10}. Among the promising candidates for qubits, each has its own distinct advantages. For instance, photons \cite{Knill01} are relatively free of the decoherence and can faithfully transmit QI between specified locations. Trapped atoms \cite{Cirac95} presents excellent coherent time typically in the range of seconds and longer. Dopants in solids \cite{Loss98,Kane98,Dutt07,Wallraff05} offer stability and potential scalability, and some of them are optics accessible and may be controlled on the order of picoseconds \cite{Press08}. A well designed quantum computer requires combining these advantages to get ``the best of two (or more) worlds'', leading to much effort devoted to investigate hybrid systems, e.g. coupling atomic systems \cite{Sorensen04,*Tian04,Andre06,*Rabl06}, quantum dots (QDs) \cite{Childress04} or nitrogen-vacancy (NV) centers \cite{Twamley10} to stripline resonators, coupling NV centers to a nanomechanical resonator \cite{Rabl09} or flux qubits \cite{Marcos10,Zhang11}, and interfaces between quantum dots and atomic systems \cite{Akopian11}.

Previous works are mainly focused on utilizing suitable physical systems at different computational steps, e.g. flying photons as transmitting medium and solid qubits as processing units. However, it is also desirable to form hybrid architectures within the same processing unit (PU), since supposing the current bottleneck problems of QDs, NV centers and other candidates were overcome, a large-scale quantum computer can be \textit{upgraded} with an extra PU, just like upgrading a classical computer with another memory bank. The extra PU can be based on a physical system even different from the original PU. In this case an architecture offering \textit{compatibility} for both physical systems demands investigation. The compatibility exhibits as follows: (1) the interface should realize coupling \textit{distant} qubits for initialization and two-qubit operations, since the extra qubits of PU are not integrated on the same chip beforehand, but upgraded afterwards and thus apart from the original ones. Theoretical achievements have been made to couple distant atoms \cite{Serafini06} or atomic ensembles \cite{Yin07}, at the same time the resonant couplings between qubits and light fields cause the system exposed to level decay and photon loss which deteriorate the accuracy of the two-qubit gate operation. To improve the case, Zheng \cite{Zheng09} suggested a scheme for two-qubit gating induced by the virtual excitation of light fields. However, (2) a major obstacle arises that different solid-state systems have distinct level structures, e.g. different emission energies, spin angular momentums, and even different configurations, thus the interface should achieve coupling \textit{nonidentical} qubits and the existing proposals coupling distant and identical qubits \cite{Serafini06,Yin07,Zheng09,Song07,YangZB11,Yang11} are infeasible for the hybrid PU. Recently, Zhang \textit{et al}. \cite{Zhang10} propose a scheme for realizing two-qubit gates with two nonresonant QDs trapped in coupled photonic crystal cavities. But the two cavities are directly coupled which does not satisfy the requirement of (1). (3) Quantum computing comprises the transmitting and readout of QI, thus the architecture should realize the state transfer from the hybrid PU to flying qubits and compatible measurement of qubits distributed over distinct physical systems. For these requirements one challenge is to transfer the state to photons independent on polarization modes, because the variation in frequencies of photons due to distinct level structures of solid qubits will bring difficulty to further process or measure the photons on polarization modes. The other challenge is to make the measurement a single-shot and furthermore a nondestructive one which is demanded for scalable quantum computation for large-scale problems \cite{Liu10}.

In this paper, to address all the above issues, we propose a novel architecture of hybird quantum processing unit (HQPU), and all the quantum computing procedures with the HQPU are controlled by coherent optical techniques. Each HQPU comprises one QD and one NV center as two solid-state qubits. The interface between qubits consists of two whispering-gallery mode (WGM) microcavities coupled to the QD and NV center respectively, and the two cavities are connected by an optical fiber. The single qubit operation is implemented by Raman process using detuned light fields, and the two-qubit operation is induced by the vacuum fields of the cavities and fiber. The proposed interface has the advantage of combining the capacity of coupling distant and nonidentical solid-state qubits and the insensitivity to population loss profiting from the virtual excitation for cavities, the fiber and solid-state qubits. The transmitting and readout components are composed of another two WGM cavities respectively coupled to the solid-state qubits. We will show that the quantum nondemolition (QND) measurement of the two-qubit state can be archived by coherent control of cavity quantum electrodynamics (QED) and optical pulses based on a proposed proposal \cite{Liu05}, and after applying the cooling processes by additional optical cycles the QI transferring to photons can be achieved.

The paper is organized as follows. In Sec. II, we describe the structure and physical realization of HQPU in detail, including the solid-state qubits, WGM microcavities and optical fiber, and give the Hamiltonian for each several part of the system. In Sec. III, we present the schemes to implement a universal set of gates including single- and two-qubit gates. The methods for readout and transmitting of QI are provided in Sec. IV and finally a summary and some prospects are made in Sec. V.

\section{Description of the HQPU architecture}
The schematic of the HQPU architecture is sketched as Fig.~1. In the subsequent subsections we describe every components of the HQPU in detail.

\begin{figure}
\includegraphics[width=8cm]{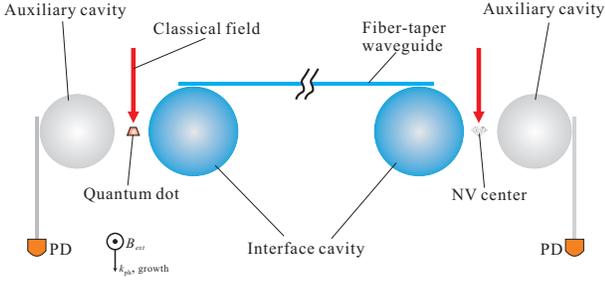}
\caption{\label{fig:1}(Color online) Schematic of the HQPU architecture. The QD and NV center are coupled to microsphere cavities (blue, ``Interface cavity'') respectively, and the two cavities are connected by a fiber-taper waveguide. A local external magnetic field is applied perpendicularly to the optical axis of the classical field and QD growth direction. Two additional microsphere cavities (grey, ``Auxiliary cavity'') are coupled to the two solid qubits respectively for the measurement and state transfer procedure. The waveguides coupled to the auxiliary cavities can lead to other destinations for further processing or directly to the photon detectors (PDs) for the readout of QI.}
\end{figure}

\subsection{Solid-state qubits of QD and NV center}
We adopt one self-assembled InAs QD as the first solid-state qubit for the HQPU architecture. The QD is formed by molecular beam epitaxy (MBE) process using the Stranski-Krastanow growth mode \cite{Petroff01}. By properly tuning the gate voltage of the $n^+$-intrinsic-Schottky (NIS) diode structure, the QD can be charged with a single electron \cite{Stinaff06}. The electron ground states $\ket{x\pm}$ serve as the logical states. In the Voigt geometry (see Fig.~1), the ground states are split by an external magnetic field (along the $x$ axis) applied perpendicularly to the QD growth $z$-direction, which is aligned parallel to the optical axis. When the QD absorbs a photon, it is excited to a trion state which consists of a singlet pair of electrons and a heavy hole. The hole spin is pinned along the growth direction due to strong confinement and spin-orbit interaction. After transforming the trion states in to the basis in the $x$-direction (denoted by $\ket{\tau+}$ and $\ket{\tau-}$), optical selection rules determine that the vertical and cross transitions couple to orthogonal linear polarizations of the optical field, leading to the four-level system shown in Fig.~2a. The QD has a typical emission energy on the order of eV (e.g. 1.39eV \cite{Greilich09}) and allows for coherent control of visible light. By applying laser pulses, the local laser-QD coupling system can be modeled by the interaction Hamiltonian (in units of $\hbar=1$) \cite{Press08}
\begin{eqnarray}
H_{\textrm{laser-QD}}=&&-\delta_e\ket{x+}\bra{x+}+\Delta_1\ket{\tau-}\bra{\tau-}+\left(\Delta_1+\delta_h\right)\ket{\tau+}\bra{\tau+}\nonumber\\
&&+\frac{1}{2}\left[\Omega_\textrm{V}(t)\left(\ket{\tau-}\bra{x-}+\ket{\tau+}\bra{x+}\right)\right.\nonumber\\
&&\left.+i\Omega_\textrm{H}(t)\left(\ket{x+}\bra{\tau-}+\ket{x-}\bra{\tau+}\right)+\textrm{H.c.}\right],\label{lqd}
\end{eqnarray}
where $\delta_e$ and $\delta_h$ are the Zeeman splittings, $\Delta_1$ is the detuning, and $\Omega_\textrm{V}(t)$ and $\Omega_\textrm{H}(t)$ are the time-dependant Rabi frequencies for the vertical and horizontal polarized components of the control pulses. Applying only vertical polarized light will reduce the system into two two-level systems. The evolution governed by Hamiltonian in Eq.(\ref{lqd}) is used to implement single-qubit operation for the QD part. The other reason for choosing this double-$\Lambda$ configuration (not the double two-level configuration based on states on $z$-direction) is relevant to the QND measurement proposal that requires selectively coupling certain transitions by different polarizations, which will be demonstrated in Sec. IV.

For the second solid-state qubit we adopt one NV center located in a diamond nanocrystal, the structure of which is distinct from the QD: each NV center is a point defect in the diamond lattice, which consists of a substitutional nitrogen atom and an adjacent vacancy \cite{Yang10}. We consider the NV center negatively charged with two unpaired electrons, and the ground states is spin triplet and labeled as $^3A$, with a splitting of $2.87$GHz between the lower level $m_s=0$ and nearly degenerated upper levels $m_s=\pm1$ \cite{Santori06}. We denote $\ket{m_s=0}=\ket{g}$ and $\ket{m_s=\pm1}=\ket{f}$ as logical states of the NV center qubit. The excited state $\ket{^3E}=\ket{e}$ is also a spin triplet. The optical transitions dictate $\ket{g}\leftrightarrow\ket{e}$ and $\ket{f}\leftrightarrow\ket{e}$ couple to $\sigma^0$ and $\sigma^+$ polarizations of light respectively \cite{Yang10apl}, forming a $\Lambda$-type system as shown in Fig.~2b. Note that the external magnetic field applied to the QD has no effect on the NV center due to the sufficiently large distance between the two qubits. The NV center has the emission energy with zero phonon line of 1.945eV \cite{Santori06} and thus also allows for optical control. By combining laser pulses, the interaction Hamiltonian governing the local laser-NV system reads
\begin{eqnarray}
H_{\textrm{laser-NV}}=\Delta_2\ket{e}\bra{e}+\left(\Omega_\textrm{V}^\prime(t)\ket{g}\bra{e}+\Omega_+(t)\ket{f}\bra{e}+\textrm{H.c.}\right).\label{lnv}
\end{eqnarray}
Here $\Omega_\textrm{V}^\prime(t)$ and $\Omega_\textrm{+}(t)$ are the Rabi frequencies for the linearly and circularly polarized pulses and $\Delta_2$ is the detuning for both pulses from the transitions under the two-photon resonance condition. This stimulated Raman process is often involved in the scheme for single-qubit gating \cite{Liu10}. The selection rules and variation in emission frequencies of the QD and NV center offers sufficient freedom for optical control and the feasibility to implement single- and two-qubit operations by different optical schemes.

\begin{figure}
\includegraphics[width=8cm]{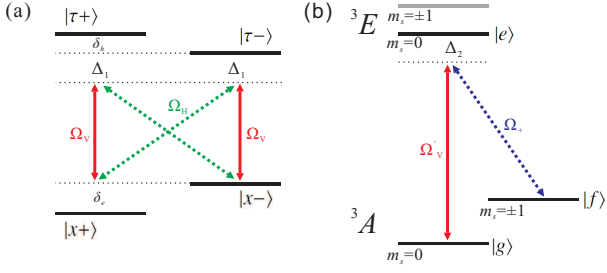}
\caption{\label{fig:2}(Color online) Energy level diagrams for the QD and NV center, induced by only local classical fields. (a) The vertical and horizontal components of the control pulse couple to the vertical (red, solid) and cross (green, dotted) transitions respectively, forming a double-$\Lambda$ system. $\delta_e$ and $\delta_h$ are the Zeeman splittings of the electron ground states and trion states, respectively, and $\Delta_1$ is the detuning. (b) Combining pulses with linearly (red, solid) and circularly (purple, dotted) polarizations, the NV center is modeled as a $\Lambda$-type system. $\Delta_2$ is the identical detuning for the two pulses. Note that to implement single-qubit operation for the NV center, the linearly pulse is replaced by the interface cavity's mode.}
\end{figure}

\subsection{The interface and components for readout and transmitting}
The interaction between the QD and NV center qubits is mediated by the interface, which consists of two microcavities connected by an optical fiber, as shown in Fig.~1. For the cavities we adopt WGM microsphere cavity, where the WGMs are induced by total internal reflection and travel along the curved boundary \cite{Vahala03}. Mircosphere cavity takes advantages of very small volume ($V_m\leq100\mu{\textrm{m}}^3$) and extremely high quality factor $Q$ ($\geq10^8$ even up to $10^{10}$) \cite{Armani03}. Moreover the fundamental WGM corresponding to the light traveling around the equator of the microsphere offers strong coupling strength between the cavity and either QD \cite{Fan99} or NV center \cite{Park06}. The solid-state qubits are individually attached and coupled to the separate microsphere cavities, namely ``interface cavities'' in the following sections. Considering only vertical polarized mode existing in the cavity, the four-level system of the QD and $\Lambda$-type system of the NV center are deduced to two two-level systems and one two-level system, respectively. Then the solid qubits and interface cavities coupling system is governed by the interaction Hamiltonian as follows
\begin{eqnarray}
H_{\textrm{qubit-cavity}}=&&g_{1}a_1e^{i\left(\Delta_1^c+\delta_e+\delta_h\right)t}\ket{\tau+}\bra{x+}+g_{1}a_1e^{i\Delta_1^ct}\ket{\tau-}\bra{x-}\nonumber\\
&&+g_{2}a_2e^{i\Delta_{2}^ct}\ket{e}\bra{g}+\textrm{H.c.},
\end{eqnarray}
where $g_{1}$ and $g_2$ are the coupling strengths between the cavity and QD and NV center respectively, $a_j$ is the annihilation operator for the $j$-th cavity mode, $\Delta_j^c$ is the detuning of the $j$-th cavity mode frequency from either QD or NV center transitions. We consider the interface cavities and solid qubits are far off resonant (large detuning $\Delta_j^c$).

For the fiber we consider an optical fibre-taper waveguide \cite{Vahala03} near the equatorial planes of both microspheres, with length $l$ and the decay rate of the cavities' fields into the continuum of the fiber modes $\tilde{\nu}$, and the number of fiber modes which significantly interact with cavity modes is of the order of $l\tilde{\nu}/2\pi{c}$. In the ``short fiber limit'' $l\tilde{\nu}/2\pi{c}\lesssim1$, only one fiber mode will essentially interact with the cavity modes \cite{Serafini06}, then the cavity-fiber coupling system can be modeled by the interaction Hamiltonian
\begin{eqnarray}
H_{\textrm{cavity-fiber}}=\nu{b}(a_1^\dag+e^{i\varphi}a_2^\dag)+\textrm{H.c.},
\end{eqnarray}
where $b$ is the annihilation operator for the fiber mode, $\nu$ is the coupling strength and the phase $\varphi$ is induced by the propagation of the field through the fiber.

For the readout and transmitting of QI, two other mircrosphere cavities (``auxiliary cavities'') are introduced and individually coupled to the qubits (see Fig.~1), meanwhile the cavities are also individually coupled to two additional optical fibre-taper waveguides, which are only for the photons escaping (output) from the cavity and then to a detector for measurements or to other destination for further processing. The auxiliary cavities and solid qubits are also off resonant.

\section{Universal quantum computing with HQPU}
\subsection{Single-qubit gate operations}
There have been experimental achievements for controlling a single QD spin \cite{Press08,Berezovsky08,Kim10} and single qubit gating of a NV center \cite{Jelezko04}, mainly using classical optical techniques. For controlling the spin of the QD, we consider the applied laser pulses with high intensity and short FWHM ($\sim4$ps), while satisfying large-detuning conditions, $\Delta_1\gg|\Omega_{\textrm{H,V}}(t)|\gg|{g_{1}}|$, which means the laser-QD interaction is much faster than the cavity-QD interaction, therefore within the pulse action one can approximatively neglect the coupling between the left cavities and QD. Then the scheme \cite{Press08} can be implanted in the HQPU architecture.

For the single-qubit gating of the NV center, we show a proposal different from the traditional laser-induced Raman process \cite{Liu10} or the scheme using microwave pulses \cite{Jelezko04}. Our proposal is based on combining the classical laser pulses ($\sigma^+$ polarized and Rabi frequency $\Omega_+(t)$) with the quantized optical field in the interface microsphere cavity (vertical polarized and coupling strength $g_2$), and the laser and the WGM field are detuned from the transitions of the NV center. This has the benefit that there is no need for the discrepancy between order of magnitudes of classical and quantized fields. The interaction Hamiltonian governing the $\Lambda$-type system can be written as
\begin{eqnarray}
H_{\textrm{1q}}=\Delta_2\ket{e}\bra{e}+\left(g_2a^\dag\ket{g}\bra{e}+\Omega_+\ket{f}\bra{e}+\textrm{H.c.}\right).\label{1qeff}
\end{eqnarray}
Under the large-detuning conditions, $\Delta_2\gg\left|\Omega_+\right|,\left|g_2\right|$, the upper level $\ket{e}$ can be adiabatically eliminated. By applying standard quantum optical techniques \cite{Gardiner91}, we obtain the effective Hamiltonian as
\begin{eqnarray}
H_{\textrm{1q}}^{\textrm{eff}^\prime}=&&\frac{\left|\Omega_+\right|^2}{\Delta_2}\ket{f}\bra{f}+\frac{\left|g_2\right|^2}{\Delta_2}a^\dag{a}\ket{g}\bra{g}\nonumber\\
&&+\frac{1}{\Delta_2}\left(g_2\Omega_+a\ket{f}\bra{g}+\textrm{H.c.}\right).
\end{eqnarray}
The first two terms of Stark shifts can be eliminated by additional pulses and initially preparing the cavity in the vacuum state, therefore the Hamiltonian Eq.~(\ref{1qeff}) can be further reduced to $H_{\textrm{1q}}^{\textrm{eff}}=g^\prime{a}\ket{f}\bra{g}+\textrm{H.c.}$, where $g^\prime=g_2\Omega_+/\Delta_2$. Under the initial condition of vacuum state, the evolution is restricted in the basis $\{\ket{f},\ket{g}\}$, and with simple calculations the evolution operator is obtain as
\begin{eqnarray}
U_\textrm{1q}\left(t\right)=\left(
  \begin{array}{cc}
    \cos\left(g^{\prime}t\right) & i\sin\left(g^{\prime}t\right) \\
    -i\sin\left(g^{\prime}t\right) & \cos\left(g^{\prime}t\right) \\
  \end{array}
\right),
\end{eqnarray}
which is used to realize the rotation of single qubit about $x$ axis. Note that the conditions of large-detuning and initially prepared vacuum fields for the single-qubit operation are in accordance with the conditions considered for the two-qubit operation in the following text.

\begin{figure}
\includegraphics[width=8cm]{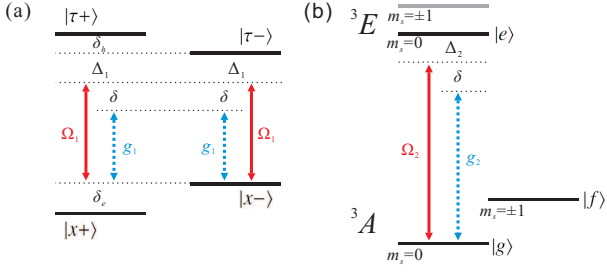}
\caption{\label{fig:3}(Color online) Energy level diagrams for the QD and NV center, induced by only vertical polarized classical fields and cavities' modes. Only the transitions $\ket{x\pm}\leftrightarrow\ket{\tau\pm}$ and $\ket{g}\leftrightarrow\ket{e}$ are enabled. $\Omega_j$ are the Rabi frequencies of the laser fields (red, solid), $g_j$ are the coupling strengths (blue, dotted) between cavities and qubits. $\delta$ is the identical difference between the detunings of laser-qubit system and cavity-qubit system.}
\end{figure}

\subsection{Two-qubit gate operation: controlled phase gate}
The major function of the interface component of the HQPU architecture is to implement two-qubit gate operation which allows two qubits of distinct physical systems to interact and work collaboratively. To attain this goal, we apply laser pulses with the vertical polarization to the QD and NV center qubits, then the cross transitions $\ket{x\pm}\leftrightarrow\ket{\tau\mp}$ and transition $\ket{f}\leftrightarrow\ket{e}$ are forbidden. Therefore the local laser-QD (Fig.~2a) and laser-NV systems (Fig.~2b) reduce to one double two-level system and one two-level system respectively, as shown in Fig.~3. Combining with the interface cavities' and fiber's modes carefully, the interaction Hamiltonian governing the whole laser-qubit-cavity-fiber system takes the form as
\begin{eqnarray}
H=H_{\textrm{laser-qubit}}+H_{\textrm{qubit-cavity}}+H_{\textrm{cavity-fiber}},
\end{eqnarray}
where
\begin{eqnarray}
H_{\textrm{laser-qubit}}=&&\Omega_1\left(e^{i(\Delta_1+\delta_e+\delta_h)t}\ket{\tau+}\bra{x+}+e^{i\Delta_1t}\ket{\tau-}\bra{x-}\right)\nonumber\\
&&+\Omega_2e^{i\Delta_2t}\ket{e}\bra{g}+\textrm{H.c.}.
\end{eqnarray}
Here $\Omega_j$ are the Rabi frequencies of the laser fields, the detunings satisfy the condition $\Delta_j^c=\Delta_j+\delta$, and the denotation $\Delta=\delta_e+\delta_h$ is adopted. We introduce three new bosonic modes $c_1, c_2$, and $c_3$, and define $a_1=\frac{1}{2}(c_1+c_2+\sqrt{2}c_0)$, $a_2=\frac{1}{2}e^{i\varphi}(c_1+c_2-\sqrt{2}c_0)$, and $b=\frac{1}{\sqrt{2}}(c_1-c_2)$, then the whole Hamiltonian in the interaction picture can be rewritten as
\begin{eqnarray}
H=&&H_0+H_i,\\
H_0=&&\sqrt{2}\nu\left(c_1^\dag{c_1}-c_2^\dag{c_2}\right),\\
H_i=&&\left[\frac{g_1}{2}\left(c_1+c_2+\sqrt{2}c_0\right)e^{i(\Delta_1+\delta+\Delta)t}+\Omega_1e^{i(\Delta_1+\Delta)t}\right]\ket{\tau+}\bra{x+}\nonumber\\
&&+\left[\frac{g_1}{2}\left(c_1+c_2+\sqrt{2}c_0\right)e^{i(\Delta_1+\delta)t}+\Omega_1e^{i\Delta_1t}\right]\ket{\tau-}\bra{x-}\nonumber\\
&&+\left[\frac{g_2}{2}e^{-i\varphi}\left(c_1+c_2-\sqrt{2}c_0\right)e^{i(\Delta_2+\delta)t}+\Omega_2e^{i\Delta_2t}\right]\ket{e}\bra{g}\nonumber\\
&&+\textrm{H.c.}.
\end{eqnarray}
We apply the unitary transformation $e^{iH_0t}$ and obtain
\begin{eqnarray}
H_I=&&\left[\frac{g_1}{2}\left(e^{-i\sqrt{2}\nu{t}}c_1+e^{i\sqrt{2}\nu{t}}c_2+\sqrt{2}c_0\right)e^{i(\Delta_1+\delta+\Delta)t}+\Omega_1e^{i(\Delta_1+\Delta)t}\right]\nonumber\\
&&\times{}\ket{\tau+}\bra{x+}+\left[\frac{g_1}{2}\left(e^{-i\sqrt{2}\nu{t}}c_1+e^{i\sqrt{2}\nu{t}}c_2+\sqrt{2}c_0\right){e}^{i(\Delta_1+\delta)t}\right.\nonumber\\
&&\left.+\Omega_1e^{i\Delta_1t}\right]\ket{\tau-}\bra{x-}+\left[\frac{g_2}{2}e^{-i\varphi}\left(e^{-i\sqrt{2}\nu{t}}c_1+e^{i\sqrt{2}\nu{t}}c_2-\sqrt{2}c_0\right)\right.\nonumber\\
&&\left.\times{}{e}^{i(\Delta_2+\delta)t}+\Omega_2e^{i\Delta_2t}\right]\ket{e}\bra{g}+\textrm{H.c.}.
\end{eqnarray}

Assuming $\Delta_j\gg|g_j|,|\Omega_j|,\sqrt{2}\nu,\delta$ and $\Delta$, the QD and NV center cannot be really excited to the upper levels by the light field and will remain in the ground states. Then the upper levels $\ket{\tau\pm}$ of QD and $\ket{e}$ of NV center can be adiabatically eliminated by the method proposed in Ref. \cite{James07}. So the first effective Hamiltonian is approximated as
\begin{widetext}
\begin{eqnarray}
H_{\textrm{2q}}^{\textrm{eff1}}=&&-\Lambda_1c_1e^{i(\delta-\sqrt{2}\nu)t}
-\Lambda_2c_2e^{i(\delta+\sqrt{2}\nu)t}
-\Lambda_0c_0e^{i\delta{t}}-\left(\kappa_{+,1}\ket{x+}\bra{x+}+\kappa_{-,1}\ket{x-}\bra{x-}-\kappa_{g,1}\ket{g}\bra{g}\right)c_2^\dag{c_1}e^{-i2\sqrt{2}\nu{t}}\nonumber\\
&&-\left(\kappa_{+,2}\ket{x+}\bra{x+}+\kappa_{-,2}\ket{x-}\bra{x-}-\kappa_{g,2}\ket{g}\bra{g}\right)c_0^\dag{c_2}e^{i\sqrt{2}\nu{t}}
-\left(\kappa_{+,0}\ket{x+}\bra{x+}+\kappa_{-,0}\ket{x-}\bra{x-}-\kappa_{g,0}\ket{g}\bra{g}\right)c_1^\dag{c_0}e^{i\sqrt{2}\nu{t}}+\textrm{H.c.}\nonumber\\
&&-\left(\varepsilon_{+,1}c_1^\dag{c_1}+\varepsilon_{+,2}c_2^\dag{c_2}+\varepsilon_{+,0}c_0^\dag{c_0}+\epsilon_+\right)\ket{x+}\bra{x+}
-\left(\varepsilon_{-,1}c_1^\dag{c_1}+\varepsilon_{-,2}c_2^\dag{c_2}+\varepsilon_{-,0}c_0^\dag{c_0}+\epsilon_-\right)\ket{x-}\bra{x-}\nonumber\\
&&-\left(\varepsilon_{g,1}c_1^\dag{c_1}+\varepsilon_{g,2}c_2^\dag{c_2}+\varepsilon_{g,0}c_0^\dag{c_0}+\epsilon_0\right)\ket{g}\bra{g},\label{eff1}
\end{eqnarray}
\end{widetext}
where $\Lambda_i$ take the forms as
\begin{eqnarray}
\Lambda_1&&=\lambda_{+,1}\ket{x+}\bra{x+}+\lambda_{-,1}\ket{x-}\bra{x-}+\lambda_{g,1}e^{-i\varphi}\ket{g}\bra{g},\nonumber\\
\Lambda_2&&=\lambda_{+,2}\ket{x+}\bra{x+}+\lambda_{-,2}\ket{x-}\bra{x-}+\lambda_{g,2}e^{-i\varphi}\ket{g}\bra{g},\nonumber\\
\Lambda_0&&=\lambda_{+,0}\ket{x+}\bra{x+}+\lambda_{-,0}\ket{x-}\bra{x-}-\lambda_{g,0}e^{-i\varphi}\ket{g}\bra{g}.
\end{eqnarray}
Here the parameters $\lambda_{\pm,i}$, $\lambda_{g,i}$, $\kappa_{\pm,i}$, $\kappa_{g,i}$, $\varepsilon_{\pm,i}$, $\varepsilon_{g,i}$, $\epsilon_\pm$, and $\epsilon_g$ are given in the appendix.

Under the conditions $\delta-\sqrt{2}\nu$, $\delta+\sqrt{2}\nu$, $\delta$, $\sqrt{2}\nu\gg\lambda_{\pm,i}$, $\lambda_{g,i}$, $\kappa_{\pm,i}$, and $\kappa_{g,i}$ $(i=0,1,2)$, the bosonic modes $c_1$, $c_2$, and $c_0$ cannot exchange energy with each other or with the classical fields \cite{Zheng09,Zhang10}. The couplings between the bosonic modes and classical fields cause energy shifts which only depend upon the numbers of QD and NV center in the ground state $\ket{x\pm}$, $\ket{g}$, while the couplings between bosonic modes lead to energy shifts depending upon both the excitation numbers of the modes and the number of QD and NV center in the ground state. The effective Hamiltonian of the second step takes the form
\begin{widetext}
\begin{eqnarray}
H_{\textrm{2q}}^{\textrm{eff2}}&&=\frac{1}{\delta-\sqrt{2}\nu}\Lambda_1\Lambda_1^\ast+\frac{1}{\delta+\sqrt{2}\nu}\Lambda_2\Lambda_2^\ast+\frac{1}{\delta}\Lambda_0\Lambda_0^\ast
+\frac{1}{2\sqrt{2}\nu}\left(\kappa_{+,1}\ket{x+}\bra{x+}+\kappa_{+,1}\ket{x-}\bra{x-}+\kappa_{+,1}\ket{g}\bra{g}\right)^2(c_2^\dag{c_2}-c_1^\dag{c_1})\nonumber\\
&&+\frac{1}{\sqrt{2}\nu}\left(\kappa_{+,2}\ket{x+}\bra{x+}+\kappa_{+,2}\ket{x-}\bra{x-}-\kappa_{+,2}\ket{g}\bra{g}\right)^2(c_1^\dag{c_1}-c_0^\dag{c_0})
+\frac{1}{\sqrt{2}\nu}\left(\kappa_{+,0}\ket{x+}\bra{x+}+\kappa_{+,0}\ket{x-}\bra{x-}-\kappa_{+,0}\ket{g}\bra{g}\right)^2(c_0^\dag{c_0}-c_2^\dag{c_2})\nonumber\\
&&-\left(\varepsilon_{+,1}c_1^\dag{c_1}+\varepsilon_{+,2}c_2^\dag{c_2}+\varepsilon_{+,0}c_0^\dag{c_0}+\epsilon_+\right)\ket{x+}\bra{x+}
-\left(\varepsilon_{-,1}c_1^\dag{c_1}+\varepsilon_{-,2}c_2^\dag{c_2}+\varepsilon_{-,0}c_0^\dag{c_0}+\epsilon_-\right)\ket{x-}\bra{x-}\nonumber\\
&&-\left(\varepsilon_{g,1}c_1^\dag{c_1}+\varepsilon_{g,2}c_2^\dag{c_2}+\varepsilon_{g,0}c_0^\dag{c_0}+\epsilon_0\right)\ket{g}\bra{g}.\label{eff2}
\end{eqnarray}
\end{widetext}
The excitation numbers of the bosonic modes are conserved during the evolution. We assume the interface cavities and optical fiber are all initially prepared in the vacuum state, therefore the bosonic modes will remain in the vacuum state during the interaction and the corresponding terms in Eq. (\ref{eff2}) can be eliminated. Then the effective Hamiltonian reduce to its final form
\begin{eqnarray}
H_{\textrm{2q}}^{\textrm{eff}}=&&\frac{1}{\delta-\sqrt{2}\nu}\Lambda_1\Lambda_1^\ast+\frac{1}{\delta+\sqrt{2}\nu}\Lambda_2\Lambda_2^\ast+\frac{1}{\delta}\Lambda_0\Lambda_0^\ast\nonumber\\
&&-\left(\epsilon_+\ket{x+}\bra{x-}+\epsilon_-\ket{x-}\bra{x-}+\epsilon_g\ket{g}\bra{g}\right).
\end{eqnarray}

The QI is encoded and distributed in the ground states of QD and NV center, presenting as a two-qubit states in basis $\{\ket{x+}\ket{g},\ket{x+}\ket{f},\ket{x-}\ket{f},\ket{x-}\ket{g}\}$. During the evolution controlled by classical fields (e.g. limited pulse duration), the solid-state qubit system undergoes an energy shift and acquires distinct phase shifts for individual basis. The evolution of the states can be written as
\begin{eqnarray}
\left\{
\begin{aligned}
\ket{x+}\ket{g}&\rightarrow{e}^{-i(\Phi_++\Phi_g+\Phi_{+g})t}\ket{x+}\ket{g},\\
\ket{x+}\ket{f}&\rightarrow{e}^{-i\Phi_+t}\ket{x+}\ket{f},\\
\ket{x-}\ket{f}&\rightarrow{e}^{-i\Phi_-t}\ket{x-}\ket{f},\\
\ket{x-}\ket{g}&\rightarrow{e}^{-i(\Phi_-+\Phi_g+\Phi_{-g})t}\ket{x-}\ket{f},
\end{aligned}
\right.\label{evolution1}
\end{eqnarray}
where
\begin{eqnarray}
\Phi_\pm=&\frac{|\lambda_{\pm,1}|^2}{\delta-\sqrt{2}\nu}+\frac{|\lambda_{\pm,2}|^2}{\delta+\sqrt{2}\nu}+\frac{|\lambda_{\pm,0}|^2}{\delta}-\epsilon_\pm,\nonumber\\
\Phi_g=&\frac{|\lambda_{g,1}|^2}{\delta-\sqrt{2}\nu}+\frac{|\lambda_{g,2}|^2}{\delta+\sqrt{2}\nu}+\frac{|\lambda_{g,0}|^2}{\delta}-\epsilon_g,\nonumber\\
\Phi_{\pm{g}}=&\frac{\lambda_{\pm,1}\lambda_{g,1}^*e^{i\varphi}}{\delta-\sqrt{2}\nu}+\frac{\lambda_{\pm,2}\lambda_{g,2}^*e^{i\varphi}}{\delta+\sqrt{2}\nu}-\frac{\lambda_{\pm,0}\lambda_{g,0}^*e^{i\varphi}}{\delta}+\textrm{c.c.}.
\end{eqnarray}
Note that the phase $\varphi$ can be eliminated by carefully choosing the length of waveguide to satisfy $\omega{l}/c=n$ experimentally. To archive a controlled phase gate, three additional single-qubit operations \cite{Kim10, Pei10} are applied subsequently
\begin{eqnarray}
\left\{
\begin{aligned}
\ket{x+}&\rightarrow{e}^{i\Phi_+t}\ket{x+},\\
\ket{x-}&\rightarrow{e}^{i\Phi_-t}\ket{x-},\\
\ket{g}&\rightarrow{e}^{i(\Phi_g+\Phi_{-g})t}\ket{g}.
\end{aligned}
\right.
\end{eqnarray}
Then the states in Eq. (\ref{evolution1}) evolve into the final form
\begin{eqnarray}
\left\{
\begin{aligned}
\ket{x+}\ket{g}&\rightarrow{e}^{-i(\Phi_{+g}-\Phi_{-g})t}\ket{x+}\ket{g},\\
\ket{x+}\ket{f}&\rightarrow\ket{x+}\ket{f},\\
\ket{x-}\ket{f}&\rightarrow\ket{x-}\ket{f},\\
\ket{x-}\ket{g}&\rightarrow\ket{x-}\ket{g}.
\end{aligned}
\right.
\end{eqnarray}
Here the solid-state qubit system undergoes a conditional phase $-(\Phi_{+g}-\Phi_{-g})t$ only if the system initially prepared in state $\ket{x-}\ket{g}$. With the choice of $(\Phi_{+g}-\Phi_{-g})t=\pi$, the controlled phase $\pi$ gate is obtained. Hence we have built a universal set of gates for optical quantum computing with the HQPU architecture, and the  collaborative work of distant QD and NV center qubits is realized. Besides, since the NV center has a relatively longer spin decoherence time than spins in QD ($T_2\sim2~\textrm{ms}$ vs. $\sim3~\mu\textrm{s}$ \cite{Ladd10}), the QI encoded in the QD can be transferred to the NV center for temporary storage, with the help of SWAP gate composed of the above demonstrated single- and two-qubit operations. Therefore the NV center can alternatively work as a ``cache memory'' in each HQPU architecture.

To analyze the efficiency and practicability of the controlled phase gate, we adopt experimentally achievable parameters \cite{Zhang10,Yang10,Yang10apl} to estimate the two-qubit interaction time for the controlled phase $\pi$ gate as an example. Using the values $\{g_1,g_2,\Omega_1,\Omega_2\}=\{0.01,0.02,0.1,0.15\}~\textrm{meV}$, and the detunings $\{\Delta_1,\Delta_2,\Delta,\nu,\delta\}=\{2,2.2,0.1,0.0145,0.02\}~\textrm{meV}$, we obtain $\{\lambda_{+,1},\lambda_{+,2},\lambda_{+,0},\lambda_{-,1},\lambda_{-,2},\lambda_{-,0},\lambda_{g,1},\lambda_{g,2},\lambda_{g,0}\}\simeq\{2.4,2.4,3.4,2.5,2.5,3.5,6.8,6.8,9.6\}\times10^{-4}~\textrm{meV}$, which satisfy the large-detuning conditions demonstrated above. Then the two-qubit interaction time for the controlled phase $\pi$ gate is estimated as $t\simeq\pi/(\Phi_{+g}-\Phi_{-g})\sim62~\textrm{ns}$, which is far less than the spin decoherence times of both QD and NV center \cite{Ladd10}. We also consider the spontaneous decay times for the excited levels of the QD as $\tau_{\textrm{QD}}\sim1.4~\textrm{ns}$ \cite{Zhang10} and NV center as $\tau_{\textrm{NV}}\sim12~\textrm{ns}$ \cite{Yang10apl}, and obtain the occupation probabilities of the excited state as $\{P_\textrm{QD},P_\textrm{NV}\}\sim\{0.3\%,0.5\%\}$ $(\simeq\Omega_i^2/\Delta_i^2)$, therefore the effective decoherence time is $t_e\sim560~\textrm{ns}$ $(\simeq\min[\tau_\textrm{QD}/P_\textrm{QD},\tau_\textrm{NV}/P_\textrm{NV}])$. So within $t_e$ many controlled phase $\pi$ gates can be implemented.

\section{QND measurements and state transfer}
The QND measurements of the qubits in the HQPU architecture can be implemented by generalizing the measurement proposal \cite{Liu05} with the help of the duad auxiliary cavities. The states of solid-state qubits undergo individual cycle and return back to the initial states, accompanied by two conditional emission of photons. The measurement cycle is constructed by a pulse sequence: firstly two resonant vertical polarized pulses are applied to the QD and NV center respectively to excite them to the corresponding upper levels, $\ket{x+}\rightarrow\ket{\tau+}$ and $\ket{g}\rightarrow\ket{e}$. Then a horizontal and a $\sigma^+$ off-resonant polarized pump pulses are adiabatically applied to the QD and NV center for tens of picoseconds and only couple the transitions $\ket{x\mp}\leftrightarrow\ket{\tau\pm}$ and $\ket{f}\leftrightarrow\ket{e}$ respectively, and the induced the Stark shifts of $\ket{\tau+}$ and $\ket{e}$ are thus resonant with the respective adjacent auxiliary cavities' modes. Within the duration of pump pulses the solid-state qubits are resonantly coupled to the respective auxiliary cavities and the excitations of QD and NV center then rapidly turns into photons by spontaneous emission and leak out into the fiber-taper waveguide. Consider the generic piece of QI encoded in the solid-state qubits is $\ket{\psi}=a\ket{x+}\ket{g}+b\ket{x+}\ket{f}+c\ket{x-}\ket{f}+d\ket{x-}\ket{g}$, then after the measurement cycles the state ideally evolves into
\begin{eqnarray}
\ket{\psi}^{\prime}=&a\ket{x+}\ket{g}\ket{1}_1\ket{1}_2+b\ket{x+}\ket{f}\ket{1}_1\ket{0}_2\nonumber\\
&+c\ket{x-}\ket{f}\ket{0}_1\ket{0}_2+d\ket{x-}\ket{g}\ket{0}_1\ket{1}_2,
\end{eqnarray}
where $\ket{1}_j$ and $\ket{0}_j$ denote the optical states depending on whether there is photon emitted from the $j$-th auxiliary cavity. The detection of the photons projects the $\ket{\psi'}$ into a basis state, which provides a QND measurement.

The state transfer from solid-state qubits to photons is based on $\ket{\psi}'$. The cooling cycles \cite{Liu10} are applied to the QD and NV center respectively, which is similar with the measurement cycle except the pump pulses are switched on preceding the resonant pulses, and the resonant pulses are horizontal and $\sigma^+$ polarized which only enable the the population transfer $\ket{x-}\rightarrow\ket{\tau+}$ and $\ket{f}\rightarrow\ket{e}$ respectively. Therefore after the cooling cycles the population are all piled on states $\ket{x+}$ and $\ket{g}$. $\ket{\psi}'$ thus transforms into
\begin{eqnarray}
\ket{\psi}^{\prime\prime}=&\ket{x+}\ket{g}\left(a\ket{1}_1\ket{1}_2+b\ket{1}_1\ket{0}_2+c\ket{0}_1\ket{0}_2+d\ket{0}_1\ket{1}_2\right),\nonumber\\
\end{eqnarray}
Hence the state is transferred to the photons propagating in the fibers and are capable to be further processed.

\begin{figure}
\includegraphics[width=8cm]{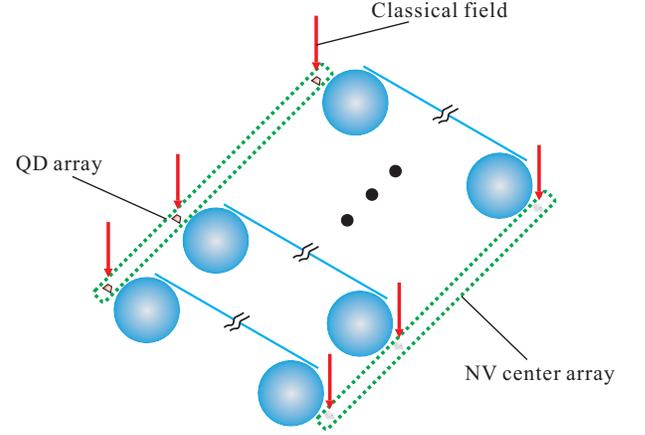}
\caption{\label{fig:4}(Color online) Schematic of the prospective hybrid quantum processor integrated by many HQPUs. The distant QDs and NV centers constitute the QD array and NV center array respectively. The nonidentical solid-state qubits within either array can be controlled by other proposals. Either array can function as a quantum processor independently, while mediated by the interface of the HQPU architecture, two arrays can work collaboratively and the quantum processor is upgraded. The components for readout and transmitting are left out.}
\end{figure}

\section{Summary and prospect}
In summary, we have depicted a blueprint of the hybrid quantum solid-state processing unit. The HQPU architecture offers compatibility for distinct physical systems, by allowing distant and nonidentical solid-state qubits to interact and work collaboratively. We also show the capability of HQPU architecture for the universal quantum computing to be all-optically controlled at every computational steps, including the initialization, single- and two-qubit operations, measurements, and state transfer. Meanwhile the virtual excitation of the solid-state qubits, cavities, and fibers guarantees the insensitivity of our methods to population loss. Moreover, the QND measurements are scalable when the number of qubits increases. Therefore, combining with the scalability of solid-state systems, the HQPU architecture is a potential key ingredient for the \textit{upgradable} large-scale quantum computer, which means, even different solid-state systems can merge and be integrated into one quantum processor afterwards. Fig.~4 shows a schematic of the prospective hybrid quantum processor integrated by many HQPUs. The distant QDs and NV centers constitute the QD array and NV center array respectively. The nonidentical solid-state qubits within either array can be controlled by other proposals (e.g. by classical fields and additional single-mode waveguides \cite{Zhang11arxiv}). Therefore either array can function as a quantum processor independently, while mediated by the interface of the HQPU architecture, two arrays can work collaboratively and the quantum processor is upgraded.

We would like to remark that other candidates for qubits may also be available provided that they are promising with stability, integratability and more importantly the optical controllability. Further research may be proceeded along these directions: one is to improve the efficiency of the QI processing within one HQPU, since the two-qubit operation is not very fast resulting from the twice adiabatical eliminations, which inherits from previous proposals based on virtual-photon. The other is, in experiments the number of fiber-taper waveguides increases with the number of the HQPU, and generally the measurements and state transfer work as parallel transmission mode \cite{Kohnen10}, which may be inconvenient for a large hybrid quantum solid-state processor. So an interesting preliminary idea is to coherently transform and encode the photon states on frequency and then combine them into a single optical fiber using a multiplexer \cite{Kim08} for transmitting or further processing.

\begin{acknowledgements}
We thank Drs. Zhang-Qi Yin, Wan-Li Yang, and Jia-Sen Jin for enlightening discussions. This work is supported by the National Nature Science Foundation of China under Grants No. 60703100, No. 10775023 and No. 11175033, and by the Fundamental Research Funds for the Central Universities under Grant No. DUT10LK10.
\end{acknowledgements}

\appendix*
\section{Parameters of the effective Hamiltonian}
The parameters of the effective Hamiltonian in Eq. (\ref{eff1}) are given as below
\begin{eqnarray}
\lambda_{+,1}&&=\frac{g_1\Omega_1^*}{4}\left(\frac{1}{\Delta_1+\delta+\Delta-\sqrt{2}\nu}+\frac{1}{\Delta_1+\Delta}\right),\nonumber\\
\lambda_{+,2}&&=\frac{g_1\Omega_1^*}{4}\left(\frac{1}{\Delta_1+\delta+\Delta+\sqrt{2}\nu}+\frac{1}{\Delta_1+\Delta}\right),\nonumber\\
\lambda_{+,0}&&=\frac{\sqrt{2}g_1\Omega_1^*}{4}\left(\frac{1}{\Delta_1+\delta+\Delta}+\frac{1}{\Delta_1+\Delta}\right),\nonumber\\
\lambda_{-,1}&&=\frac{g_1\Omega_1^*}{4}\left(\frac{1}{\Delta_1+\delta-\sqrt{2}\nu}+\frac{1}{\Delta_1}\right),\nonumber\\
\lambda_{-,2}&&=\frac{g_1\Omega_1^*}{4}\left(\frac{1}{\Delta_1+\delta+\sqrt{2}\nu}+\frac{1}{\Delta_1}\right),\nonumber\\
\lambda_{-,0}&&=\frac{\sqrt{2}g_1\Omega_1^*}{4}\left(\frac{1}{\Delta_1+\delta}+\frac{1}{\Delta_1}\right),\nonumber\\
\lambda_{g,1}&&=\frac{g_2\Omega_2^*}{4}\left(\frac{1}{\Delta_2+\delta-\sqrt{2}\nu}+\frac{1}{\Delta_2}\right),\nonumber\\
\lambda_{g,2}&&=\frac{g_2\Omega_2^*}{4}\left(\frac{1}{\Delta_2+\delta+\sqrt{2}\nu}+\frac{1}{\Delta_2}\right),\nonumber\\
\lambda_{g,0}&&=\frac{\sqrt{2}g_2\Omega_2^*}{4}\left(\frac{1}{\Delta_2+\delta}+\frac{1}{\Delta_2}\right),\nonumber
\end{eqnarray}
\begin{eqnarray}
\kappa_{+,1}&&=\frac{\left|g_1\right|^2}{8}\left(\frac{1}{\Delta_1+\delta+\Delta-\sqrt{2}\nu}+\frac{1}{\Delta_1+\delta+\Delta+\sqrt{2}\nu}\right),\nonumber\\
\kappa_{+,2}&&=\frac{\sqrt{2}\left|g_1\right|^2}{8}\left(\frac{1}{\Delta_1+\delta+\Delta+\sqrt{2}\nu}+\frac{1}{\Delta_1+\delta+\Delta}\right),\nonumber\\
\kappa_{+,0}&&=\frac{\sqrt{2}\left|g_1\right|^2}{8}\left(\frac{1}{\Delta_1+\delta+\Delta-\sqrt{2}\nu}+\frac{1}{\Delta_1+\delta+\Delta}\right),\nonumber
\end{eqnarray}
\begin{eqnarray}
\kappa_{-,1}&&=\frac{\left|g_1\right|^2}{8}\left(\frac{1}{\Delta_1+\delta-\sqrt{2}\nu}+\frac{1}{\Delta_1+\delta+\sqrt{2}\nu}\right),\nonumber\\
\kappa_{-,2}&&=\frac{\sqrt{2}\left|g_1\right|^2}{8}\left(\frac{1}{\Delta_1+\delta+\sqrt{2}\nu}+\frac{1}{\Delta_1+\delta}\right),\nonumber\\
\kappa_{-,0}&&=\frac{\sqrt{2}\left|g_1\right|^2}{8}\left(\frac{1}{\Delta_1+\delta-\sqrt{2}\nu}+\frac{1}{\Delta_1+\delta}\right),\nonumber\\
\kappa_{g,1}&&=\frac{\left|g_2\right|^2}{8}\left(\frac{1}{\Delta_2+\delta-\sqrt{2}\nu}+\frac{1}{\Delta_2+\delta+\sqrt{2}\nu}\right),\nonumber\\
\kappa_{g,2}&&=\frac{\sqrt{2}\left|g_2\right|^2}{8}\left(\frac{1}{\Delta_2+\delta+\sqrt{2}\nu}+\frac{1}{\Delta_2+\delta}\right),\nonumber\\
\kappa_{g,0}&&=\frac{\sqrt{2}\left|g_2\right|^2}{8}\left(\frac{1}{\Delta_2+\delta-\sqrt{2}\nu}+\frac{1}{\Delta_2+\delta}\right),\nonumber
\end{eqnarray}
\begin{eqnarray}
\varepsilon_{+,1}&&=\frac{|g_1|^2}{4(\Delta_1+\delta+\Delta-\sqrt{2}\nu)},\varepsilon_{+,2}=\frac{|g_1|^2}{4(\Delta_1+\delta+\Delta+\sqrt{2}\nu)},\nonumber\\
\varepsilon_{+,0}&&=\frac{|g_1|^2}{2(\Delta_1+\delta+\Delta)},\varepsilon_{-,1}=\frac{|g_1|^2}{4(\Delta_1+\delta-\sqrt{2}\nu)},\nonumber\\
\varepsilon_{-,2}&&=\frac{|g_1|^2}{2(\Delta_1+\delta+\sqrt{2}\nu))},\varepsilon_{-,0}=\frac{|g_1|^2}{2(\Delta_1+\delta)},\nonumber\\
\varepsilon_{g,1}&&=\frac{|g_2|^2}{4(\Delta_1+\delta-\sqrt{2}\nu)},\varepsilon_{g,2}=\frac{|g_2|^2}{2(\Delta_2+\delta+\sqrt{2}\nu)},\nonumber\\
\varepsilon_{g,0}&&=\frac{|g_2|^2}{2(\Delta_1+\delta)},\epsilon_+=\frac{|\Omega_1|^2}{2(\Delta_1+\Delta)},\epsilon_-=\frac{|\Omega_1|^2}{2\Delta_1},\nonumber\\
\textrm{and}\nonumber\\
\epsilon_0&&=\frac{|\Omega_2|^2}{2\Delta_2}.
\end{eqnarray}

\end{document}